# Exploring Disorder in the Spin Gapless Semiconductor Mn$_2$CoAl

Robert G. Buckley[1,2], Tane Butler[1], Catherine Pot[1], Nicholas M. Strickland[1] and Simon Granville[1,2]

[1]Robinson Research Institute, Victoria University of Wellington, Lower Hutt 5046, New Zealand
[2]MacDiarmid Institute for Advanced Materials and Nanotechnology, New Zealand

**Abstract**
Since the prediction of spin-gapless semiconducting behaviour in the Heusler compound Mn$_2$CoAl, evidence of spin-gapless behaviour in thin films has typically been inferred from magnetotransport measurements. The spin gapless state is however fragile, and further, band structure calculations indicate that even a small amount of atomic disorder may destroy it. To explore the impact of disorder on the properties of Mn$_2$CoAl, we have undertaken an experimental study of the structural, magnetotransport and optical properties from the far infrared to the UV, on DC magnetron sputtered Mn$_2$CoAl thin films. A very short mean free path, of the order of a lattice spacing, is extracted from the DC transport data. A room temperature resistivity of 200 µΩ.cm along with a small and negative temperature coefficient of resistance between 4 and 400 K was measured. We note that parameters of this magnitude are often observed in disordered metals. We find this behaviour is well described by a weak localisation model, a result that is supported by a large Drude contribution to the optical response, where a high scattering rate is derived, which is equal to the value derived from the DC conductivity and Hall effect data. We also note the strong similarities between the magnetotransport behaviour reported for Mn$_2$CoAl films in the literature, including ours. We conclude that, based on comparisons between the experimental data, and recent band structure calculations that explicitly include disorder, as-prepared Mn$_2$CoAl films are best described as a disordered metal, rather than a spin gapless semiconductor.

## I. Introduction

Spintronic devices are of great interest because they can potentially resolve the emerging limitations in silicon-based technologies. For this endeavour to be successful however, there is an urgent need to understand better the properties of recently identified new spintronic materials. Ideally, they need to meet a number of technical requirements including 100% spin polarisation, high electronic carrier mobility, low saturation magnetic moment, and compatibility with existing semiconductor technologies. [1] One extensively studied spintronic-materials subclass are the half-metals; these exhibit metallic conductivity for one spin polarisation and a gap in the density of states at the Fermi level for the other spin polarisation. Band structure calculations on the ideal inverse Heusler structure have shown that another subclass of compounds, spin gapless semiconductor (SGS) [1–8], can also exist; these materials would exhibit, instead of half-metallicity, an energy gap of zero width for one spin polarisation and a gap in the density of states at the Fermi level for the other spin polarisation. The calculations indicated that these compounds meet several of the key technical requirements noted above, including 100% electron and hole spin-polarisation, and they exhibit a very low threshold energy for carrier excitation, making their spin polarisation potentially voltage-tunable.

Based on band structure calculations, the inverse Heusler compound $Mn_2CoAl$ has been identified as a SGS. [3-8] The key findings of the band structure calculations are that there is strong hybridization between the d-electrons of the transition metal atoms, with a contribution from the Al p-electrons, resulting in dispersed bonding and antibonding bands. The majority-spin bands display an indirect zero-width gap between the Brillouin zone centre at the $\Gamma$ and the X point and a direct gap of about 0.8 eV at $\Gamma$. The minority-spin bands display a direct gap at $\Gamma$ of about 0.3 eV. This band structure is consistent with the mobility of the carriers being higher than in many semiconductors and both the electrons and holes can be 100% spin polarised as required for spintronic devices. Kudrnovsky and co-workers [9] have, however, pointed out that predictions of the features described above, including the SGS state, depend not only on the computational method employed and the parameters required by the calculation, but also on preserving the ideal stoichiometry and composition. In particular, the degree and type of antisite disorder is of major significance. They observe that with increasing antisite disorder the gap of zero-width is lost as electronic states fill the majority-spin 'gap' between the valence and conduction bands, creating a "pseudo-gap", although the half-metallic nature remains. Their calculations, along with those of Galanakis et al, [10] show that disorder drives a merging of the usual crystalline distinct and sharp density of state features, and that the net magnetic moment, the anomalous Hall effect, and the DC conductivity can all be significantly affected by antisite disorder.

In the first experimental report on arc-melted bulk samples of Mn2CoAl, Ouardi and co-workers experimentally identified $Mn_2CoAl$ as a spin gapless semiconductor based on the measured near temperature-independent conductivity, the linear magnetoresistance, a vanishing Seebeck coefficient and a relatively small anomalous Hall conductivity. [5] Experimental studies have also been undertaken on thin films of $Mn_2CoAl$ and interpreted in terms of the SGS state being robust, although in most cases varying degrees of anti-site and compositional disorder have been identified [5,11–19]. A notable feature common to most of these studies on thin films is that the expected superlattice XRD reflections associated with an ideal ordered inverse Heusler structure are generally not observed, although there are exceptions.[4,5,11] For Heusler compounds, with the standard formula $X_2YZ$, composed of elements with similar atomic numbers it is often difficult to distinguish even the two related structural types, Heusler ($L2_1$) or inverse Heusler (XA) that differ in their X/Y ordering, as they both exhibit the same superlattice lines.[19,20] Moreover, Heusler compounds are well known to exhibit antisite disorder, and depending on the degree of disorder, the resulting structures can either be B2 (X/Y disorder) or A2 (complete disorder).[11,19,20] The commonly used experimental probe, XRD, has difficulty distinguishing between these structures.[20] It is generally concluded that the lack of superlattice lines, as appears to be the case in most reports, indicates the presence of disorder in $Mn_2CoAl$, although it is difficult to be confident that the ideal inverse Heusler structure has been formed based on XRD measurements alone.

In several reports the saturation magnetisation at low temperature is less than the calculated value of $2\mu_B$ per formula unit.[11,14,16,17] Calculations of the total magnetic moment indicate that this parameter is also strongly dependent on the degree and type of anti-site and compositional disorder, further implying that the films exhibit variable disorder.[9,10] The saturation magnetisation is clearly a more sensitive quantitative measure of intersite disorder than XRD.

Interestingly, the reported DC magnetotransport results for all samples are similar – we have collected these for all available published studies in Table I. In both bulk and film samples, the observed room temperature DC resistivities and the temperature coefficients of resistance between 4 and 300 K are remarkably consistent, with similar magnitudes of approximately 100 to 700 $\mu\Omega$-cm and approximately $-2 \times 10^{-7}$ $\Omega$-cm. K$^{-1}$ respectively. Carrier densities range from $10^{20}$ to $4 \times 10^{22}$ cm$^{-3}$. The calculated mobilities are comparable to those seen in silicon, for similar carrier densities at 300 K, however they do not demonstrate any systematic behaviour with carrier density.[21]

Although several authors have drawn attention to the presence of anti-site disorder in their samples, the impact of such disorder has generally not been invoked in understanding the physical properties of $Mn_2CoAl$ and thus the robustness of the SGS state has not been fully established.

| $Mn_2CoAl$ preparation technique | $\rho$ (300 K) ($\Omega$-cm) | $\dfrac{d\rho}{dT}$ ($\Omega$-cm.K$^{-1}$) | $n_e$ (cm$^{-3}$) | $\mu$ (cm$^2$/V.s) | $M_S$ ($\mu_B$/f.u.) | Ref. |
|---|---|---|---|---|---|---|
| **Arc-melt bulk** | $4.1 \times 10^{-4}$ | $-1.4 \times 10^{-7}$ | $3 \times 10^{20}$ (300 K) | 51 | 2 | 5 |
| **Induction-melt bulk** | $2.56 \times 10^{-4}$ | non-linear | - | - | 0.56 | 17 |
| **Molecular beam epitaxy film** | $2.9 \times 10^{-4}$ | non-linear | $4 \times 10^{22}$ (300 K) | 0.5 | 1 | 11 |
| **Sputter film** | $0.87 \times 10^{-4}$ | non-linear | $1.6 \times 10^{20}$ (4 K) | 0.45 | 1.95 | 13 |
| **Sputter film** | $2.7 \times 10^{-4}$ | $-1.2 \times 10^{-7}$ | $2 \times 10^{21}$ (4 K) | 11.6 | 1 | 14 |
| Ar ion assisted sputter film | 6.9 x 10$^{-4}$ | -4.8 x 10$^{-7}$ | 6 x 10$^{20}$ (4 K) | 3.2 | 2 | 15 |
| Molecular beam epitaxy film | 3.75 x 10$^{-4}$ | -1.55 x 10$^{-7}$ | ~10$^{22}$ (10 K) | ~1.5 | 1.4 | 16 |
| Molecular beam epitaxy film | 3.6 x 10$^{-4}$ | -2.2 x 10$^{-7}$ | - | - | - | 18 |
| Sputter film | 2.0 x 10$^{-4}$ | -0.7 x 10$^{-7}$ | 2 x 10$^{22}$ (10 K) | 2 | 1.02 | Current sample |

Table I: Magnetotransport and magnetisation derived parameters for Mn$_2$CoAl samples from the literature, where ρ is resistivity, $n_e$ is carrier concentration, μ is mobility and $M_S$ is saturation magnetization. The carrier density is observed to be only weakly temperature dependent.

The aim of our study is to develop a better understanding of the role disorder plays in determining the physical properties of Mn$_2$CoAl. To that end we report a study of the resistivity, magnetoresistance, Hall effect including the anomalous Hall effect, and IR/visible reflection spectroscopy of sputtered thin films of Mn$_2$CoAl. Based on XRD, EDAX, and magnetisation measurements we conclude that our films do exhibit antisite and compositional disorder. We observe a DC resistivity at ambient of 200 μΩ.cm along with a small and negative temperature coefficient of resistance between 4 and 400 K (Table I). We find that the mean free path is of the order of 1 nm and is largely independent of temperature. Further, the scattering rates derived from the DC transport data and the spectroscopic study are similar at ~10$^{14}$ per second. Given that the DC transport behaviour of both bulk and thin film samples is similar, exhibiting a small, negative and approximately linear temperature dependent resistivity, we conclude that the DC transport behaviour of Mn$_2$CoAl can only be understood if disorder is explicitly accounted for and it is best described by weak localisation in a disordered metal. Thus, and as indicated by band structure calculations incorporating disorder, Mn$_2$CoAl as prepared to date is best described as exhibiting a disordered metallic-like behaviour.

II. Experimental Section

Thin films of Mn$_2$CoAl were deposited by DC magnetron sputtering using a Kurt J Lesker CMS-18 system, from a Torus 2" source with a high magnet strength assembly. The voltage applied during the film growth was 326-340 V for a sputtering current of 290-310 mA resulting in a growth rate of 0.7-0.8 Å.s$^{-1}$. The vacuum chamber base pressure was 2 x 10$^{-8}$ Torr and the films were sputtered onto 10 x 10 mm$^2$ MgO[001] substrates from a composite target of Mn:Co:Al with a relative composition of 50:25:25. The substrates were cleaned in acetone, isopropyl alcohol, and deionised water before drying in a flow of nitrogen gas, then loaded into the vacuum chamber and additionally cleaned for 5 minutes with an Argon plasma. The substrates were held for 1 hour at the deposition temperature before film deposition was started. The films were sputtered under 5 to 6 mTorr of Ar to thicknesses of 100 to 150 nm at a growth rate of 0.83 Å.s$^{-1}$. Dektak profilometry was employed to determine the film thicknesses. After growth, the films were cooled to room temperature at a rate of ~10 K/minute. Deposition temperatures between 450 °C and 550

°C were used, and as the structural, magnetic, magneto-transport and optical behaviour were observed to be similar for each temperature, we discuss only data from films grown at 550 °C.

XRD spectra were collected using a Bruker D8 Advance system with Co $K_\alpha$ X-ray radiation at a wavelength of 1.79 Å. Using the EDAX capability of a FEI Quanta SEM the film composition was determined to be $Mn_{1.7}Co_{1.2}Al_{1.1}$ The compositional analysis calculation employed CalcZAF [22] and NIST DTSA-II [23] software and EDAX measurements from elemental standards.

Temperature- and field- dependent resistivity and Hall effect measurements were made in a Quantum Design PPMS with the resistivity option. A Quantum Design MPMS-7 SQUID magnetometer was employed to make magnetization measurements.

The reflectivity of multiple $Mn_2CoAl$ films were measured at room temperature and at 77 K employing a Bruker V80v interferometer from 50 $cm^{-1}$ (6.2 meV) to 35,000 $cm^{-1}$ (~4.3 eV) using a variety of light sources, beam splitters and detectors. The reflectivity was measured relative to a freshly evaporated Al film and was corrected for the aluminium reflectivity using literature values [24]. The uncertainty in the magnitude of the reflectivity was estimated to be about 2% based the reproducibility of the measurement and the mismatch associated with a change in resources. The spectra of all samples were observed to be temperature independent.

### III. Results and Discussion

#### A. Structural characterisation

$Mn_2CoAl$ consists of four face-centred cubic sublattices, two of Mn and one each of Co and Al, formed in the XA $Hg_2CuTi$-type or inverse Heusler structure with space group 216 ($F\bar{4}3m$). Each sublattice, in the order Mn-Mn-Co-Al, is equally spaced along the [111] direction and follows the Wyckoff site sequence, 4a-4c-4b-4d [25]. Figure 1 shows a θ-2θ XRD scan of a sputtered film, where the out of-plane (002) and (004) reflections of $Mn_2CoAl$ are clear. The presence of these peaks from films deposited on MgO[001] substrates indicates that the films grow aligned with the substrate structure. The c-axis lattice constant derived from this measurement is 5.74 Å which is a little smaller than the value of ~5.80-5.86 Å reported previously for textured thin films grown at temperatures below 400 °C [8,12,13,15,16] or for bulk $Mn_2CoAl$ [5,17,26], but is very close to the c-axis constant found for films grown at 500 °C [14]. The in-plane lattice parameter, determined from the (220) peak of the film, is a=5.78 Å (not shown). The c/a ratio is 1.008, showing the film grown at 550 °C is very close to the expected cubic structure, and the cell volume of 191.76 $Å^3$ is 1.7% smaller than in the bulk [5]. No evidence was found for the superlattice (111) reflection, which would be present if the film were in either the ideal inverse Heusler XA or the ideal regular Heusler $L2_1$ ($Cu_2MnAl$-type) phase. This data is consistent with disorder between the Mn(4c) and Al(4d) sites as discussed in detail by Liu et al [4]. We note that of the thin film reports, only Jamer et al. reported the presence of superlattice peaks [11], otherwise they were absent [13,14,18], or not mentioned [15,16].

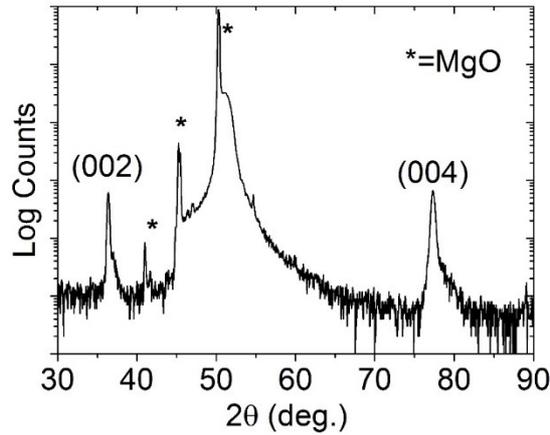

**Figure 1:** A θ-2θ XRD scan for a film grown at 550 °C where the out of-plane (002) and (004) reflections of $Mn_2CoAl$ are clear. The MgO [00$\ell$] lines are indicated by an asterisk, *, showing that the film's growth is aligned with the substrate structure. There is no evidence of the ideal inverse Heusler XA ($Hg_2CuTi$-type) or the ideal regular Heusler $L2_1$ structure-related superlattice lines. Note that Co $K_\alpha$ X-ray radiation at a wavelength of 1.79 Å was used.

### B. Magnetisation and magnetotransport

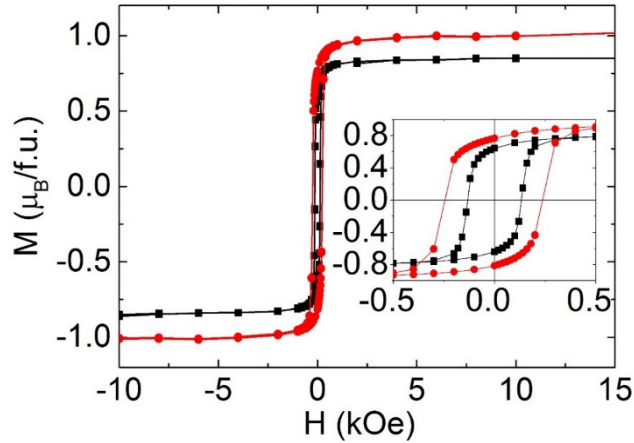

**Figure 2:** Magnetisation vs field loops of a film grown at 550 °C, measured with an in-plane field at 300 K (squares) and 10 K (circles). The saturation magnetisation at 10 K is 1.02 $\mu_B$ per unit cell and the coercive field approximately 250 Oe. The inset shows an expanded view of the same data around zero field.

Figure 2 displays typical magnetisation hysteresis loops for a $Mn_2CoAl$ film collected at 10 K and 300 K. The film has a 10 K saturation magnetisation of 1.02 $\mu_B$ per unit cell, equivalent to 195 emu/cc, which is above the 300 K value by 15%. This is approximately half the expected saturation of 2 $\mu_B$ per unit cell [4,9,10] although as noted in Table I there are reports on films displaying a similar saturation magnetisation near 1 $\mu_B$.

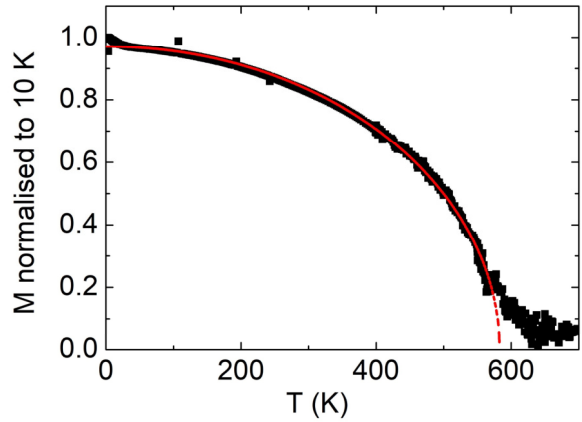

**Figure 3:** Temperature-dependent magnetisation shows that the $T_c$ of the $Mn_2CoAl$ film is approximately 585 K. The red dashed line is an extrapolation to estimate the Curie temperature as described in the text. The applied magnetic field was 1 kOe.

The temperature dependent magnetisation was measured from 3 to 400 K with the SQUID magnetometer, and from 300 to 700 K using the VSM oven of the PPMS, and the combined data are shown in Figure 3. The data deviate from the Bloch $T^{3/2}$ law and fit the empirical relation $M(T) = M(0)[1 - (T/T_C)^2]^{1/2}$ (red dashed line). Extrapolating this fit to zero magnetisation gives a Curie temperature of approximately 585 K, in agreement with other thin film results [13] but smaller than the 720 K of bulk $Mn_2CoAl$ [5].

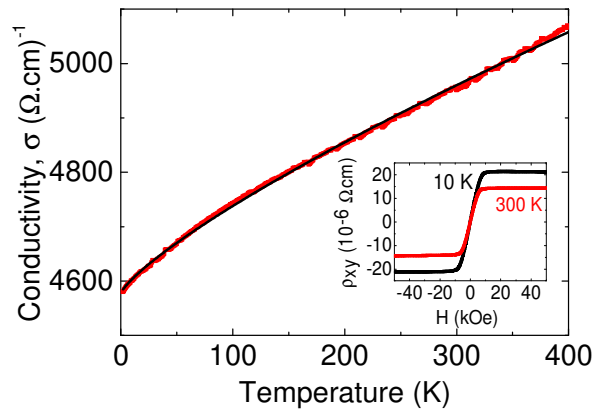

**Figure 4:** The temperature-dependent (4-400 K) DC conductivity (red points) and a fit to a weak localisation model (black line) discussed in the text. Inset: anomalous Hall resistivity at 300 K and 10 K.

Displayed in Figure 4 is the DC conductivity from 4 to 400 K. The conductivity is relatively low at 5000 (Ω-cm)$^{-1}$ at 300 K, and interestingly exhibits a negative, and approximately linear, temperature coefficient of resistance of about -0.7 x 10$^{-7}$ Ω-cm/K, although on close inspection there is a small downward curvature below 100 K. As shown in Table I, for both the bulk and thin film samples, the magnitude of the 300 K resistivity in general exhibits a remarkably similar 300 K magnitude of approximately 100-700 µΩ-cm. Moreover, there is little difference in reported values of the temperature coefficient of resistance (TCR) in the range between 4 and 300 K. The TCR is observed to be small and negative, approximately -2 x 10$^{-7}$ Ω-cm/K (see Table I). Such a negative, small, and approximately linear TCR in either a pure metal or a semiconductor is unusual, although commonly observed in disordered metals. [27]

The inset in Figure 4 displays the field dependent Hall resistivity at 10 K and 300 K. The observed anomalous component of the Hall resistivity ranges from 10 to 20 µΩ.cm between 300 and 10 K and is similar to reported values. Calculations by Kudrnovsky et al. [9] show that Mn(4c)-Al(4d) antisite disorder will increase the anomalous Hall resistivity from the expected value of zero at 0 K. Arima et al. [16] have also concluded that their anomalous Hall measurements on $Mn_2CoAl$ can be understood in terms of disorder.

Although the magnitude of the conductivity makes estimating the carrier concentration from the Hall effect uncertain, for our films we get a temperature independent value of approximately $2\times10^{22}$ electrons per $cm^3$. When combined with the measured conductivity we get a mean free path of under 1 nm and a Hall mobility of ~2 $cm^2$/V.s over the full range of temperatures measured. Using literature data (Table I) the calculated mean free paths of other researchers differ by less than a factor of five for all $Mn_2CoAl$ samples. Similarly, the calculated Hall mobility varies between 2 to 50 $cm^2$/V.s for most samples. The short mean free paths, the relatively low mobility, and their weak dependence on temperature are consistent with the presence of significant disorder.

The significant similarities of the magnetisation, structure and transport of the $Mn_2CoAl$ samples reported by us and in the literature to date indicate that this material has been synthesised more-or-less reproducibly by several different groups. However, the general lack of superlattice reflections in XRD, the lower magnetisation found by several groups, and the differences in the mobility are reasons to question this conclusion. As identified by band structure calculations, the presumption of a spin gapless electronic structure relies on maintaining the ideal stoichiometry and atomic order. Kudrnovsky et al. [9] showed that atomic swaps between the Mn(4c) and Al(4d) sites close the gap in the majority spin channel, eliminating the spin gapless nature of the electronic structure, so that it becomes a half-metal. They also showed that the anomalous Hall conductivity increases with the level of Mn(4c)/Al(4d) disorder. Galanakis et al. [10] calculated both the electronic structure and the atomic-resolved magnetic moments for several possible atomic swaps and found that in all cases the majority spin channel has closed already at 5% atomic swaps. The closing of the majority spin gap was also found by Xin et al [26] in the case of atomic swaps, and by Feng et al. [6] for Mn-deficient $Mn_2CoAl$.

A reduced magnetic moment can be readily explained by considering deviations from the ideal composition, which is not often explicitly reported in the $Mn_2CoAl$ literature. Taking the experimental composition of our Mn-deficient films, $Mn_{1.7}Co_{1.2}Al_{1.1}$, the atomic-resolved magnetic moments from Galanakis et al [10] of Mn(4a) = -1.55 $\mu_B$, Mn(4c) = +2.70 $\mu_B$, Co(4c)= 0.21 $\mu_B$, Al(4c) = -0.08 $\mu_B$, Co(4b) = +0.89 $\mu_B$ and Al(4d) = -0.05 $\mu_B$, and assuming that the excess Co and Al sits on the Mn(4c) sites, we calculate an expected magnetisation of 1.214 $\mu_B$. The experimentally measured saturation magnetisation of ~1.02 $\mu_B$/f.u. is reached with only an additional 4% atomic swaps between the 4c (Mn) and 4b (Co) sites.

To highlight the importance of atomic disorder to the magnetotransport, we analyse the magnetotransport properties of the films in terms of weak localisation, a framework developed for disordered metals. [27,28] Mooij [29] has identified a correlation for disordered metallic alloys between the resistivity and the TCR with a low and negative TCR and a resistivity greater than approximately 150 µΩ-cm. Under these conditions, the mean free path is short, the electronic transport is qualitatively described as diffusive rather than ballistic, and quantum interference occurs between incoming and scattered waves, resulting in an enhanced backscattering and thus enhanced resistivity. At low temperature, elastic scattering dominates, and phase coherence of the carrier wave function is maintained, but as the temperature increases inelastic scattering increases, destroying the phase coherence, and the resistivity decreases.

Dugdale gives the three-dimensional correction to the Boltzmann conductivity due to weak localisation as [27]:

$$\Delta\sigma(T) = \left(\frac{e^2}{2\pi^2\hbar}\right)(D\tau_{in})^{-1/2},$$

where D is the diffusion coefficient and $\tau_{in}^{-1}$ is the inelastic scattering rate. This correction reduces the conductivity and can be significant when the electron mean free path is short, of the order of the electron wavelength. As the temperature rises, inelastic scattering increases, for example due to phonons, the correction diminishes due to loss of phase coherence, and the conductivity increases as we observe in Figure 4. The temperature dependence of the correction is given by the temperature dependence of the inelastic scattering rate, $\tau_{in}^{-1}$, and is generally written as

$$\tau_{in}^{-1}(T) = \beta T^p,$$

where β and $p$ depend on the inelastic scattering mechanism. The solid black line in Figure 4 is a fit of the correction term to the measured conductivity assuming that the normal Boltzmann contribution is only weakly temperature dependent [31,32]. We find a good fit and that $p$ is 1.6 and β/D equals 2.5 x $10^{15}$ $K^{-p}s^{-1}$. Assuming a diffusion coefficient of 5 x $10^{-5}$ $m^2s^{-1}$ then β = 1.25 x $10^{11}$ $K^{-p}s^{-1}$. It needs to be recognised, however, that there are potentially several inelastic scattering mechanisms. [28,30-33] For example, if the inelastic scattering is phonon dominated then in the clean limit, where the electron mean free path is long compared to the phonon wavelength, $p$ will be one. On the other hand, in the dirty limit values of 2 or 3 have been observed for phonon scattering. [33,34] Other inelastic scattering mechanisms are of course possible. [28] A key conclusion however, is that the good fit by the expression above to our conductivity data, from 4 to 400 K, indicates a single dominating scattering mechanism.

The approach described above applies to non-magnetic disordered alloys. Dugaev et al [35] have extended the weak localisation correction to ferromagnetic disordered alloys including spin-orbit scattering and show that the correction can be observed where the elastic mean free path is very short resulting in a negative magnetoresistance at all fields. As noted above we observe a mean free path of less than 1 nm so as-deposited films are clearly in this limit. Following Dugaev et al. [35] it can be shown that the internal magnetic induction required to eliminate a weak localisation correction in this limit is some hundreds of Tesla, so well above the observed internal magnetic induction in $Mn_2CoAl$. To illustrate this point, we show in Figure 5 the magnetoresistance of $Mn_2CoAl$ measured between 2 and 300 K. Magnetic fields up to 50 kOe are applied in the plane of the sample. The magnetoresistance is field-dependent and weakly temperature-dependent and is negative at all temperatures with a sharp negative dip centred at zero field. The largest change in the magnitude of the magnetoresistance is small, at -0.22%, and increases slightly less than linear with increasing applied magnetic field. The magnitudes and negative behaviour are similar to that reported by other authors [11.13,14,18] on thin $Mn_2CoAl$ films, although with the exception of Chen et al. [18] they do not report the monotonic decreasing magnetoresistance with decreasing temperature observed in the inset of Figure 5. The unusual negative, small, and linear magnetoresistance reported in $Mn_2CoAl$ films is consistent with a weak localisation correction to the conductivity of a ferromagnet, in the dirty limit, in which spin-orbit scattering is present. [35] This result is very different from that reported from an arc-melt bulk sample [5] where at temperatures of 100 K and below the magnetoresistance is positive. Dugaev et al. [35] also emphasis that the sign of the magnetoresistance depends on the interplay between the magnitudes of the magnetisation and the spin-orbit scattering rate.

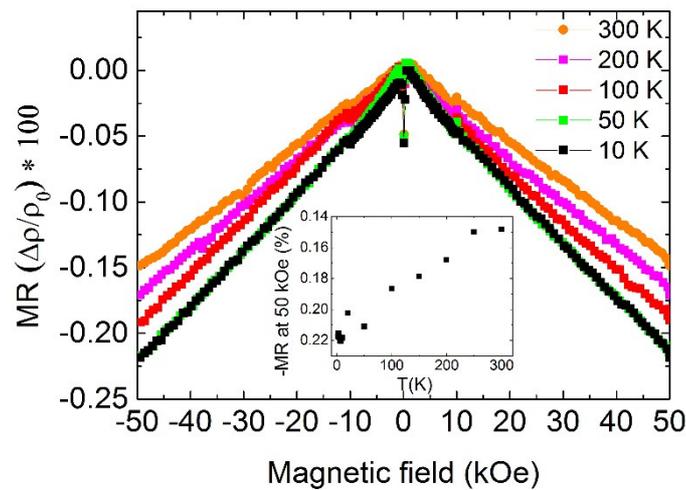

**Figure 5.** Magnetoresistance measured with magnetic field parallel to the film plane. Inset: the 50 kOe absolute magnetoresistance variation with temperature.

### C. Reflectivity measurements

To gain further insight into the physical properties of $Mn_2CoAl$, and in particular the impact of disorder, we have probed the optical response of our films by measuring the room temperature reflectivity between 100 cm$^{-1}$ and 35,000 cm$^{-1}$ as illustrated in Figure 6. The background infrared reflectivity is relatively high at ~78% and from 8,000 cm$^{-1}$ it falls in value to ~18% by 35,000 cm$^{-1}$. The reflectivity peaks at 263 cm$^{-1}$ (33 meV), 319 cm$^{-1}$ (40 meV) and 396 cm$^{-1}$ (49 meV) we assign to phonons in the film and the substrate.

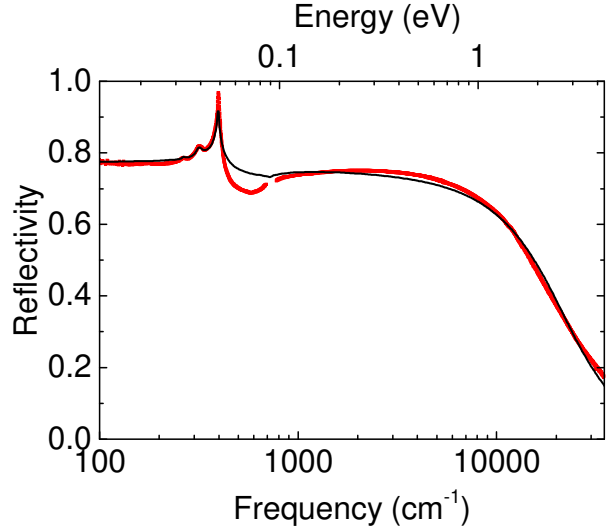

**Figure 6.** The measured reflectivity (red dots) of Mn$_2$CoAl plotted against frequency on a semi-log scale from 100 to 35,000 cm$^{-1}$. As described in the text the continuous black line is a two-layer reflectivity model that includes for the Mn$_2$CoAl film: a Drude term; three Lorentz oscillators to account for the phonons; and one Lorentz oscillator to describe the high frequency interband transitions. For the MgO substrate the measured reflectivity was fit with a series of Lorentz oscillators giving a dielectric function that compared well with the literature. [36,37]

To extract a frequency dependent dielectric function from the reflectivity of our Mn$_2$CoAl film we have used public domain software, RefFIT, to identify the 'best-fit' dielectric function describing the reflectivity [38]. As our relatively low conductivity Mn$_2$CoAl film is 100 nm thick it is semi-transparent so the strong phonon absorption in the MgO substrate contributes to the observed optical response near 400 cm$^{-1}$. This required the employment of a two-layered optical model; a 100 nm Mn$_2$CoAl film on a 0.05 cm thick MgO substrate. Based on separate reflectivity measurements, model dielectric functions were developed first for the MgO substrate alone, and second for the film-substrate system. These models consist of a series of Drude-Lorentz oscillators as described by the equation below for the frequency dependent complex dielectric function:

$$\varepsilon(\omega) = \varepsilon_\infty + \sum_i \frac{S_i^2}{\omega_{0i}^2 - \omega^2 - i\gamma_i\omega} \ .$$

Where $\varepsilon_\infty$ is the high frequency dielectric constant, $\omega_0$ is the oscillator frequency, S is the oscillator strength, and *i* denotes the oscillator in question. $\gamma$ is the scattering rate when applied to a Drude term ($\omega_0$) or damping when applied to a vibrational or electronic mode. The derived MgO frequency dependent dielectric function compared very well with literature reports. [36,37] This substrate model was then incorporated into a two-layered film-substrate model to develop a dielectric function of the semi-transparent Mn$_2$CoAl film.

Displayed in Figure 6 (black solid line) is the 'best-fit' two-layer reflectivity model compared to a measured reflectivity spectrum, one of three films deposited at 550 °C that were measured and modelled. The derived dielectric function for Mn$_2$CoAl consists of a Drude or 'free-carrier' term ($\omega_0 = 0$), three Lorentz oscillators to describe the phonons, and a Lorentz oscillator to account for high frequency electronic transitions. The best-fit Drude-Lorentz parameters are listed in Table II below. The fit is reasonable, given the simplicity of the model, except at frequencies near 500 cm$^{-1}$ associated with the strong substrate phonon. We demonstrate below that the poorer fit in this spectral region does not negatively impact our conclusion. The derived parameters for the dielectric function varied by less than 50% from sample-to-sample.

| Drude-Lorentz Fitting Parameters | | | |
|---|---|---|---|
| Model (i=1 to 5) | $\omega_0$ (cm$^{-1}$) | S (cm$^{-1}$) | $\gamma$ (cm$^{-1}$) |
| $\varepsilon_\infty$ = 1.0 | | | |
| Free Carriers (Drude) | 0 | 25,000 | 2,986 |
| Phonon | 265 | 265 | 9 |
| Phonon | 316 | 1,009 | 26 |
| Phonon | 387 | 1,209 | 32 |
| Electronic transition | 33,265 | 19,264 | 42,683 |

**Table II**. A list of the best-fit Drude-Lorentz parameters for Mn$_2$CoAl. The parameters are defined in the equation for the dielectric function above.

In Figure 7 we plot the real part of the conductivity, $\sigma_1$, ($\sigma(\omega)=i\omega\varepsilon(\omega)$, where $\varepsilon = \varepsilon_1 + i\varepsilon_2$ is the complex dielectric function and $i$ is the imaginary unit) derived from the reflectivity model. To demonstrate the dominating role of the 'free-carriers' across the entire spectral region, we have also plotted just the Drude term (dashed black line). The conductivity extrapolates to a zero-frequency value of 3500 ($\Omega$-cm)$^{-1}$, remarkably close to the measured DC value of 5000 ($\Omega$-cm)$^{-1}$. Moreover, the Drude parameters, the plasma frequency (25,000 cm$^{-1}$) and the scattering rate (~ 9x10$^{13}$ sec$^{-1}$), are also close to the corresponding values derived from the DC conductivity and Hall effect data i.e. 42,000 cm$^{-1}$ and 2 x 10$^{14}$ sec$^{-1}$ respectively. It is notable, then, that a description of the infrared behaviour requires a high scattering rate similar to that derived from the DC transport data. Combining the DC and optical observations emphases that as-prepared Mn$_2$CoAl films are best described as disordered metals exhibiting weak localisation. To complete the picture the real ($\varepsilon_1$) and imaginary ($\varepsilon_2$) parts of the complex dielectric function are also displayed in the inset of Figure 7.

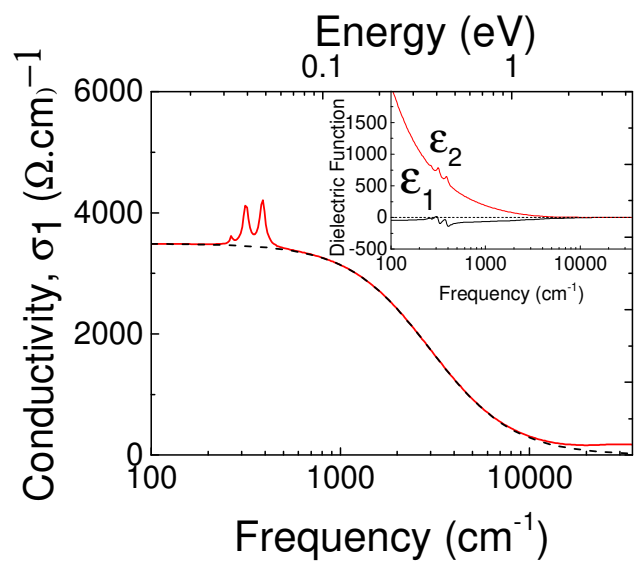

**Figure 7.** The real part of the frequency dependent conductivity (red line) for $Mn_2CoAl$ derived from the reflectivity by fitting a number of oscillators using RefFIT [38]. The black dashed line is the Drude component of the model and illustrates the dominant role played by the strongly scattering free carrier / Drude response. Note that the 300 K DC conductivity is 5000 $(\Omega\text{-cm})^{-1}$ a reasonable match to the AC value at 100 $cm^{-1}$ of 3500 $cm^{-1}$. Inset: the real and imaginary parts of the complex dielectric function.

The DC and frequency dependent conductivity data presented here compares well with data from Shreder et al.[17] in the overlapping spectra range above 0.1 eV, i.e. in both cases $\sigma_1$ is around 3500 $(\Omega\text{-cm})^{-1}$. Shreder et al. [17] do report however, that $\varepsilon_1$ is positive and rising towards low frequencies. While we report a negative $\varepsilon_1$, it is only just negative, which is driven by the high scattering rate of the Drude response, and $\varepsilon_1$ goes positive in a limited frequency range due to the relatively strong phonons around 40 meV, below their measured range.

Band structure calculations show that with increasing intersite and compositional disorder, the majority spin channel states are added in the region of the zero-width gap, thus removing the spin gapless semiconductor nature of $Mn_2CoAl$. [9,10]. In the calculations a gap however, remains in the minority spin channel i.e. defected $Mn_2CoAl$ films are half-metals. The calculations also show a broadening and merging of the features usually observed in the crystalline density of states. At high frequencies, we find that an additional oscillator centred at about 30,000 $cm^{-1}$ (~4 eV) is required to model the measured reflectivity. This oscillator is very broad and presumably accounts for the expected low energy direct interband transitions of $Mn_2CoAl$ although it is at a higher energy than predicted in published band structure calculations. Although this feature is at the edge of the range of reliable data the general conclusion is that the band structure calculations are in broad agreement with the observed spectral measurements.

$\varepsilon_2$ exhibits phonons at 265 $cm^{-1}$ (33 meV), 316 $cm^{-1}$ (39 meV) and 387 $cm^{-1}$ (48 meV). It needs to be emphasised that the high frequency phonon lies very close to the phonon feature associated with the MgO substrate for which we don't have a perfect model so the assignment has to be considered tentative. Using the Correlation Method [39] we found that the irreducible representation for the space group $F(\bar{4}3m)$ (#216), consists of 3 infrared active modes of $F_2$ symmetry, and equal to the number observed. This is also consistent with calculated phonon dispersion curves where it is predicted there are 3 pairs of degenerate zone-centre vibrations at 246, 260 and 362 $cm^{-1}$ in reasonable agreement with the observed modes at 265 $cm^{-1}$, 316 $cm^{-1}$ and 387 $cm^{-1}$ respectively [6,8,40].

### IV.    Conclusions

We have carried out a comprehensive experimental study of $Mn_2CoAl$ thin films while explicitly considering throughout the effect of composition and intersite disorder on the observed electronic properties. We find that XRD, EDAX, and magnetisation measurements indicate that our films possess compositional and antisite disorder. The films are Mn-deficient with a composition determined to be $Mn_{1.7}Co_{1.2}Al_{1.1}$. XRD and magnetisation measurements indicated that the films also exhibit antisite disorder. The 300 K DC resistivity is observed to be 200 $\mu\Omega.cm$ with a small and negative TCR down to 4 K. We also find the mean free path is of the order of 1 nm. A single Drude contribution with a large scattering rate of approximately $10^{14}$ per second largely accounts for the optical response and a similar scattering rate is derived from the DC conductivity. These values are typically observed in disordered metals and we find the conductivity can be well described by a weak localisation model between 4 and 400 K. These results are thus consistent with band structure calculations in which antisite and compositional disorder has been included. The calculations indicate that, as disorder is introduced, states are added to the majority channel in the region of the zero-width band gap, while the minority channel keeps a gap so that $Mn_2CoAl$ remains a half-metal. We find that the magnetotransport properties of most $Mn_2CoAl$ films reported in the literature display remarkably similar behaviour (see Table I) and can also be better understood as disordered metals rather than spin gapless semiconductors.

In summary, the observed electronic properties of our as-prepared $Mn_2CoAl$ films are consistent with band structure calculations that explicitly incorporate intersite and compositional disorder. The films can be understood in terms of a disordered metal and quantifiably described by a weak localisation model, indicating that as-prepared $Mn_2CoAl$ films are best understood as disordered metals rather than spin gapless semiconductors.

While addressing issues raised by the referees we became aware of a paper that has arrived at a similar conclusion based on an in-depth structural study. [41]

**Acknowledgements**

The authors acknowledge valuable discussions with Joe Trodahl, and Barney Tennyson for EDX analysis.